\documentstyle[aps,prl,preprint]{revtex}
\begin{document}
\title{Magnetic Phases of the 2D Hubbard Model at Low Doping}
\author{Andrey V. Chubukov \and Karen A. Musaelian}
\address{Department of Physics, University of Wisconsin - Madison,
1150 University Avenue, Madison, WI 53706}
\date{\today}
\maketitle
\begin{abstract}
We study the equilibrium spin configuration of the
2D Hubbard model at low doping, when a long-range magnetic order is still
present. We use the spin-density-wave formalism and identify three different
low-doping regimes depending on the value of $z = 4U \chi_{2D}$ where
$\chi_{2D}$ is the Pauli susceptibility of holes.
 When $z < 1$, the collinear
 antiferromagnetic state remains stable upon low doping.
As candidates for the ground state for $z > 1$ we first examine
the planar spiral phases with the pitch $Q$ either in one or in both
spatial directions. Mean-field calculations favor the spiral $(\pi,Q)$ phase
for $1 < z < 2$, and $(Q,Q)$ phase for $z > 2$.
  Analysis of the bosonic modes of the spiral state  shows that the
$(Q,Q)$ state has a negative longitudinal stiffness and is unstable towards
domain wall formation. For the $(\pi,Q)$ state, the longitudinal stiffness is
positive, but to the lowest order
in the hole concentration, there is a degeneracy between this state
and a whole set of noncoplanar states.
These noncoplanar states are characterized by two order
parameters, one associated with a spiral,
 and the other with a commensurate
 antiferromagnetic ordering in the direction perpendicular to the plane of a
spiral.
 We show that in the next order in
the hole concentration this degeneracy is lifted,
favoring noncoplanar states over
the spiral. The equilibrium, noncoplanar configuration is found to be close to
the N\'{e}el state with a small
spiral component whose amplitude is
proportional to the square root of
the hole concentration. These findings lead to a novel scenario of spin
reorientation upon doping in Hubbard antiferromagnets.
\end{abstract}
\narrowtext

\section{introduction}
\label{intro}

Magnetic properties of the $CuO_2$ layers in
high temperature superconductors have been recently
attracting an intense interest
as magnetism is possibly a major contributor to
 the  mechanism of superconductivity~\cite{magnetism}.
There are numerous reasons to believe that the magnetic
 properties of weakly doped cuprates are quantitatively
 captured by the effective 2D theory for one degree of freedom per $CuO_2$ unit
which is provided by a one-band Hubbard model
\begin{equation}
{\cal H} = -\sum_{i,j} t_{i,j} a^{\dagger}_{i,\alpha} a_{j,\alpha} + U \sum_{i}
n_{i,\uparrow} n_{i,\downarrow}
\label{hamilt}
\end{equation}
Here $\alpha$ is a spin index, $n = a^{\dagger} a$, and
$t_{i,j}$ is a hopping integral which acts mainly between nearest ($t$) and
next-nearest ($t^{\prime}$) neighbors.  We will assume that
$t^{\prime}$ is negative~\cite{hyber}.

At half-filling, the ground state of the 2D
Hubbard model exhibits a long range commensurate
N\'{e}el order provided $t^{\prime}$ is not very large.
It has been known for many years~\cite{and,khomskii}
that holes introduced
into a commensurate antiferromagnet give rise to a long-range dipolar
distortion of the staggered magnetization. In 2D this effect was studied in
detail by  Shraiman and Siggia~\cite{siggia}. They found
that in the simplest scenario, the dipolar distortion
 leads to a spiral spin configuration
with the momentum $(\pi,Q)$ at any nonzero doping. The
incommensurate $(\pi,Q)$ phase was also obtained in the early
perturbative studies of the Hubbard model with small $U$~\cite{Schultz},
 and in several other
mean-field~\cite{john,lee,dombre} and self-consistent~\cite{subir}
calculations.

In this paper, we use the spin-density-wave (SDW) approach and study the
structure of magnetic correlations in the Hubbard model at small but finite
doping when long-range magnetic order is still present. We will show
that depending on the strength of the interaction between the holes, three
different solutions of the Hubbard model at low doping
 are possible: (i) commensurate Neel phase at small interaction
(precise criteria will be derived below), (ii) phase separation at sufficiently
strong interaction, and (iii) intermediate
homogeneous incommensurate phase  which, however, differs from the planar
spiral
suggested by the mean-field analysis. This new incommensurate phase
is noncoplanar in the spin space, and its properties are
much closer to the properties of a commensurate $(\pi,\pi)$
spin configuration than those of a planar spiral.

The analysis presented here is related to other works on incommensurate
magnetic phases at finite doping. Previous mean-field studies of the
Hubbard and $t-J$ models~\cite{siggia,john,lee,dombre,piers} have focused on
the three configurations with ordering momenta $(\pi,\pi)$, $(\pi,Q)$
and $(Q,Q)$, and have
shown that
in a certain range of parameters, any of these
configurations  can have energy lower than that of the other two states.
Our energy analysis  is consistent with their results.
These mean-field studies also found
 that for short-range repulsion,
the $(Q,Q)$ phase is likely to be unstable towards domain walls
formation~\cite{siggia,dombre}, while the $(\pi,Q)$ phase is stable.
Shraiman and Siggia~\cite{boris}
 developed a macroscopic theory of the bosonic
excitations in the $(\pi,Q)$
spiral phase.
Surprisingly,  to the lowest order in density
they found a peculiar degeneracy in
 the ground state energy for the planar spiral state
and for a whole set of
noncoplanar magnetic configurations with the plane of the spiral varying in
space. This degeneracy is also present in our  microscopic calculations
reported below.
 However, the further assumption of Shraiman and Siggia that
the next-order terms in doping concentration  stabilize
the planar spiral
state is inconsistent with our microscopic calculations which favor noncoplanar
spin configuration for the same range of parameters as in ~\cite{boris}.

The paper is organized as follows. In Sec~\ref{m-f} we consider the mean-field
theory of the spiral phases in the Hubbard model and compare the ground state
energies of different phases. In Sec~\ref{stab} we compute the dispersion of
the bosonic excitations in the $(\pi,Q)$ and $(Q,Q)$ spiral phases to the
lowest order in hole density and find
that the $(Q,Q)$ spiral has negative longitudinal susceptibility,
 while the
$(\pi,Q)$ spiral has an infinite number of zero modes. We will identify the set
of magnetic states which are degenerate in energy with the planar $(\pi,Q)$
spiral. In the next section,
we show that the degeneracy is lifted by the next-to-leading order terms in
hole density. We compute the ground state energy and find the equilibrium
spin configuration at a finite doping. We then discuss the properties of this
equilibrium state, in particular, the behavior of the dynamical spin
susceptibility. Finally, Sec~\ref{concl} states our conclusions.

\section{the spiral phases}
\label{m-f}

In this section, we describe spin-density wave (SDW) calculations for the
Hubbard model at and near half-filling. Let us start with the commensurate
N\'{e}el
state. The SDW approach for this state
has been discussed
several times in the literature~\cite{john,SWZ,tes,bedell,chub_fren}
and we will use the results of these studies.
At half-filling, the fermionic spectrum consists of
conduction and valence bands separated by the energy gap
$\Delta = U \langle S_z\rangle$.
The dispersion relation for the valence fermions is
$E^{d}_k = - E^- + \epsilon^+$ where $E_- =\sqrt{\Delta^2 + (\epsilon^-)^2}$
and $\epsilon^- = -2t(\cos k_x + \cos k_y)$,
 $\epsilon^+ = 4|t^{\prime} \cos k_x \cos k_y|$.
 This dispersion has
a maximum at four points $(\pm\pi/2,\pm\pi/2)$ in the center of each of the
edges of the magnetic Brillouin zone,
 provided that $t^{\prime}$ is not too large.
In the neighborhood of these points, $E^{d}_k$ can be presented as
$E^{d}_k=-\Delta + p_{\parallel}^{2}/2m_{\parallel} +
p_{\perp}^{2}/2m_{\perp}$.
Thus near $(\pi/2,\pi/2)$, we have
 $p_{\parallel}=(k_x - k_y)/2, ~p_{\perp}=(k_x + k_y)/2$), $m_{\parallel} =
(4t^{\prime})^{-1}$, $m_{\perp} = (4J-4|t^{\prime}|)^{-1}$ where
$J =4t^2/U$ is the inverse bandwidth.  For both $La-$ and
$Y-$based materials, $t^{\prime} \sim J$ and therefore both masses scale as
$1/J$. Note, that the minimum at $(\pi/2,\pi/2)$ is rather robust --- even if
$t^{\prime} =0$ and the mean-field spectrum is degenerate along the whole edge
of the magnetic Brillouin zone, the actual dispersion still has four minima at
$(\pm\pi/2,\pm\pi/2)$ due to quantum fluctuations
(not included in our present SDW treatment)~\cite{vignale,Manous}.

At finite doping, the chemical potential moves into a valence band, and the
states near the maximum of $E^{d}_k$ become empty. These states are often
referred to as hole pockets.
For a commensurate spin ordering,
 all four pockets
become equally occupied.  Performing then simple
 calculations, we obtain that in the
presence of holes, the ground state energy of the $(\pi,\pi)$ phase changes to
\begin{equation}
E^{(\pi,\pi)} = \frac{\pi x^2}{4\sqrt{m_\perp m_\parallel}}
\label{epipi}
\end{equation}

We will also need the expression for the
static magnetic susceptibility of the $(\pi,\pi)$ state.
In the SDW formalism, the susceptibility  is obtained by
summing up the series of bubble
diagrams, and the result is
\begin{equation}
{\overline\chi}^{+-} (q) = \frac{\chi (q)}{ 1 - U\chi (q)}
\label{chitot}
\end{equation}
where for $\Delta >t, t^{\prime}$
\begin{eqnarray}
\chi^{+-}(q) &=&
 \frac{1}{N} ~{\sum_{E^{d}_k <\mu}}^{\prime}
 \left[1 -
\frac{\epsilon^{-}_k \epsilon^{-}_{k+q} - \Delta^2}{E^{-}_k
E^{-}_{k+q}}\right]~\frac{1}{E^{c}_k -
 E^{d}_{k+q}}
 \nonumber \\
&&  \frac{1}{N} ~{\sum_{\stackrel{E^{d}_{k+q} <\mu}{E^{d}_{k} >\mu}}}^{\prime}
 \left[1 +
\frac{\epsilon^{-}_k \epsilon^{-}_{k+q} - \Delta^2}{E^{-}_k
E^{-}_{k+q}}\right] \frac{1}{E^{d}_{k} - E^{d}_{k+q}}
\label{chi0}
\end{eqnarray}
Prime at the summation signs indicate that the summation is over the magnetic
Brillouin zone.
Near ${\bar Q} = (\pi,\pi)$, the
static transverse susceptibility should
 obey the  a hydrodynamic relation~\cite{hydro}
$~\chi^{+-}_{\rm st}(q) =
 2N_0^2 /(\rho_s(q-{\bar Q})^2)$,
where ${\bf N}_0$ is the on-site magnetization, and
$\rho_s$ is the spin-stiffness. At half-filling
only bubbles containing valence and conduction fermions are
allowed, and performing simple calculations,
we obtain $\rho_s = \rho^{0}_s = J( 1 - 2(t^{\prime}/t)^2)/4$,
where $J = 4t^2/U$. At finite doping, there is
also a contribution to the stiffness from the last term in (\ref{chi0})
which contains only valence
fermions.
This last contribution is proportional
 to the Pauli susceptibility,
which in two dimensions does not depend on carrier concentration.
As a result, in the  SDW approximation,
 the  spin stiffness acquires a finite correction
at an arbitrarily small deviation from
half-filling~\cite{siggia,dombre,chub_fren}
\begin{equation}
\rho_s = \rho^{0}_s (1-z),
\end{equation}
where
\begin{equation}
z = 4U\chi_{\rm 2D}^{\rm pauli} =
2U\frac{\sqrt{m_{\perp}m_{\parallel}}}{\pi}
\end{equation}
We see that if $z<1$, the Neel state remains stable at finite doping, while if
$z <1$, it becomes unstable, and we have to consider incommensurate states as
possible candidates to the ground state.
In the naive mean-field calculations
discussed so far, we have $z \sim U/J \gg 1$, which implies that
 commensurate state becomes unstable immediately upon doping.
However, we have explicitly verified
that all our conclusions  are valid for arbitrary ratios of $t/U$ and
$t^{\prime}/t$ provided that magnetic order at half-filling is commensurate,
and the hole pockets are located at $(\pm \pi/2,\pm \pi/2)$.
We therefore will consider $z$ as an imput parameter which can, in principle,
have any value. Some rigorous
results about the dispersion of bosonic excitations at arbitrary
$t,t^{\prime},U$ will be presented in appendix B.

It is worth emphasizing that even in the large $U$ limit,
the value of $z$ can in fact be of the order of unity. The point is
that the mean-field results at large $U$
must be taken with caution in view of strong self-energy and
vertex corrections which both contribute powers of
$U/tS$. Self-consistent consideration of these corrections indicate
that they do not change the momentum dependence of the vertices at small
$q-{\bar Q}$, but reduce the overall scale of the
effective interaction between holes from $U$
to $U_{eff}$ which is of the order of the bandwidth $J$~
\cite{siggia,chub_fren,klr}. This in turn implies that
 $z$ is in fact simply a number, independent of $U/t$.
Furthermore, striktly speaking, instead of $U_{\rm eff}$
we have to consider the total
 scattering amplitude of two holes, $T$. In two-dimensions,
\begin{equation}
T=\frac{U_{\rm eff}}{1+ \frac{mU_{\rm eff}}{4\pi}\log
\frac{mU_{\rm eff} p_{\rm F}}{4\pi}}~~,
\end{equation}
Hence $T$, and
 consequently, $z$ vanishes logarithmically as
the hole concentration, $\delta \propto \sqrt p_F$, tends to zero.
This always makes the collinear antiferromagnetic
state stable at {\em very} low doping~\cite{chub_fren}.
However, the range of doping where logarithmic corrections
are important is likely to be very small,
 and in this paper we simply set $T=U_{\rm eff}$
and  consider $z = O(1)$ as a doping-independent parameter.
For simplicity, throughout the paper, we will focus on  large $U$  and
present the results for the quantities, not related to Pauli
susceptibility, only to the leading order in $t/U$.

We now consider incommensurate spin configurations at finite doping.
Let us first focus on the two simplest candidates:
 the spiral states with the ordering vectors $(\pi, Q)$ and
$(Q,Q)$~\cite{siggia,dombre,subir,piers}. As before, we will label the ordering
momentum as ${\bar Q}$ (${\bar Q}$ can be either $(\pi,Q)$ or $(Q,Q)$).
The spin order parameter now has two components, and
in terms of the fermionic operators is expressed as
\begin{eqnarray}
&S^+_{\bar Q} &=\sum_k\langle a^{\dag}_{k,\uparrow}a_{k+\bar Q,\downarrow}
\rangle \equiv  S_{\bar Q}~,\nonumber\\
&S^-_{-\bar Q} &= \sum_k\langle a^{\dag}_{k,\downarrow}a_{k-\bar Q,\uparrow}
\rangle = (S_{\bar Q})^*~.
\label{a1}
\end{eqnarray}
For definiteness, we choose the order parameter to be in the XY plane.
Without a loss of generality one can also choose $S_{\bar Q}$ to be real.
The real space spin configuration described by (\ref{a1})
is then $S^X_R = S_{\bar Q}\cos({\bf \bar QR})$,
and $S^Y_R = S_{\bar Q}\sin({\bf \bar QR})$,
where $\bf R$ denotes the lattice site.

The SDW calculations proceed in the same way as before: one has to decouple
the interaction term in (\ref{hamilt}) using (\ref{a1}) and
diagonalize the quadratic form.
 Performing the computations, we obtain
\begin{eqnarray}
E_k^{c,d} = \epsilon_+ \pm E_- ,
\label{spir_ener}
\end{eqnarray}
\label{a2}
where as before $E_- =\sqrt{\Delta^2 + (\epsilon_-)^2}$, but now
\begin{equation}
\epsilon_+ = \frac{\epsilon_{k+\bar Q/2} + \epsilon_{k-\bar Q/2}}{2}~,
\epsilon_- = \frac{\epsilon_{k+\bar Q/2} - \epsilon_{k-\bar Q/2}}{2}~.
\end{equation}
 Note that for XY-ordering,
these new electron states appear as hybridization
of the original electrons with opposite spins and thus
contain no spin labels.
In this situation, there is no doubling of the unit cell,
and the summation over momenta in (\ref{a2}) is extended to the whole first
Brillouin zone $(-\pi/a < k_{x,y} < \pi/a)$.

The gap $\Delta$ is related to the parameters of the Hubbard model
via the self-consistency condition
\begin{equation}
\frac{1}{U} = \frac{1}{N}\sum_k
\frac {1}{2\sqrt{\Delta^2+E_-^2}}
\label{self-conf}
\end{equation}
where the summation goes over the momenta of occupied states.

Consider now specifically the spiral $(\pi,Q)$ state.
The mean-field fermionic spectrum for this state is not symmetric with
respect to the reflection $k_y \rightarrow -k_y$, although it is still
symmetric with respect to $k_x \rightarrow -k_x$. Consequently,
the minima at $(\pm \pi/2, \pi/2)$ have lower hole energy than those
at $(\pm\pi/2,-\pi/2)$.

Suppose that the concentration of holes filling the lower energy pocket
is $x_1$, and the higher energy pocket --- $x_2 ~(x_1 + x_2 =x)$.
 Simple calculations then show
that the ground state energy of the
 $(\pi,Q)$ phase  is given by
\begin{equation}
E^{(\pi,Q)} = -t \bar q(x_1 - x_2) +
\frac{t^2{\bar q}^2}{4\Delta} +
\frac{\pi(x^2_1 + x^2_2)}{2\sqrt{m_\perp m_\parallel}}~,
\end{equation}
where ${\bar q} \equiv {\bar q}_y = \pi -Q$.
The first term reflects the decrease in total energy due to the $\epsilon_+$
term in the spectrum, the
 second term results from the redistribution of the energy levels below
the Fermi level, and the last term corresponds to the increase of the total
 energy due to unequal occupation of pockets.
 Minimizing the total energy of the $(\pi,Q)$ state
with respect to $\bar q$ we obtain to the lowest order in hole concentration
\begin{equation}
\bar q = \frac{U}{t}(x_1 - x_2) .
\end{equation}
For the energy of the $(\pi,Q)$ state, we then have
\begin{equation}
E^{(\pi,Q)} = \frac{\pi (x_1-x_2)^2}{4\sqrt{m_\perp m_\parallel}} (1 - z) +
\frac{\pi x^2}{4\sqrt{m_\perp m_\parallel}} .
\label{eqpi}
\end{equation}
We see that the ground state energy of the $(\pi,Q)$ phase becomes
smaller than that of the $(\pi,\pi)$ phase at $z >1$,
exactly when the stiffness for the $(\pi,\pi)$ phase becomes negative.
In the latter case, we also have
 $x_1 = x$ and $x_2 = 0$, which implies that
only  two out of four pockets are occupied, and $\bar q = (U/t)x$. A similar
analysis was performed in~\cite{dombre}.

Finally, consider the $(Q,Q)$ phase. Now the fermionic
spectrum is
not symmetric with respect to either the $k_x \rightarrow -k_x$ or
$k_y \rightarrow -k_y$ reflections. The point $(\pi/2,\pi/2)$ becomes
the only absolute minimum of the hole spectrum. Suppose that the concentration
of holes filling the lowest energy pocket is $x_1$, the two intermediate energy
pockets - $x_2$, and the highest energy pocket - $x_3$
($x_1 + 2 x_2 + x_3 =x$). Then the ground state
energy of the $(Q,Q)$ phase is given by
\begin{equation}
E^{(Q,Q)} = -2t\bar q(x_1 - x_3) + \frac{t^2{\bar q}^2}{2\Delta} +
\frac{\pi(x_1^2+2x_2^2+x_3^2)}{\sqrt{m_\perp m_\parallel}}.
\end{equation}
The inverse pitch of the spiral ${\bar q} = \pi - Q$ is found by minimizing
the energy:
\begin{equation}
\bar q = \frac{U}{t}(x_1-x_3).
\end{equation}
The total energy then assumes the following form:
\begin{equation}
E^{(Q,Q)} = \frac{\pi}{2\sqrt{m_\perp m_\parallel}}\left((x_1-x_3)^2(1-z) +
(x-2x_2)^2+4x_2^2\right) .
\end{equation}
It is immediately obvious, that if $z<1$, then the lowest possible energy is
achieved when $x_1=x_2=x_3$, and $\bar q = 0$,
i.e. in the $(\pi,\pi)$ phase. If $z>1$, then $x_3 = 0$. Minimizing the energy
with respect to $x_1$, we find
\begin{equation}
x_1 = \frac{4}{6-z}x
\end{equation}
when $z < 2$, and
\begin{equation}
x_1 = x
\end{equation}
when $z > 2$. In the former case only the highest energy pocket has no holes,
and the total energy is equal to
\begin{equation}
E^{(Q,Q)} = \frac{2\pi x^2}{\sqrt{m_\perp m_\parallel}}\frac{2-z}{6-z} .
\end{equation}
It is straightforward to see that it is always higher than the energy of
the $(\pi,Q)$ phase, given by eq.(\ref{eqpi}).
In case of $z > 2$, only the pocket with the lowest hole energy is occupied,
and the total energy is
\begin{equation}
E^{(Q,Q)} = \frac{\pi x^2}{2\sqrt{m_\perp m_\parallel}}(2-z) .
\label{eqq}
\end{equation}
Comparing (\ref{epipi}) (\ref{eqpi}) and (\ref{eqq}), we observe that
the N\'{e}el state is the minimum for $z<1$, the
 $(\pi,Q)$ spiral phase has the lowest energy
at $1<z<2$, while  the $(Q,Q)$ state
has the lowest energy at $z>2$.

Observe however, that in the latter case,
 $\partial^2 E^{(Q,Q)}/\partial x^2 \sim (2-z)$
is negative. On general grounds, this result
 suggests that the homogeneous solution is unstable.
We will see in the next section that the longitudinal stiffness for the $(Q,Q)$
state is in fact negative - this will be  another argument in favor of an
inhomogeneous ground state. On the other hand, for the
$(\pi,Q)$ phase in its region of stability ($1<z<2$) we have,
$\partial E^{(\pi,Q)}/\partial x^2 >0$, i.e., homogeneous solution is stable.

\section{Collective excitations}
\label{stab}

Now we proceed to examining the stability of the spiral states by considering
collective bosonic excitations.
It follows from general considerations that in a spiral phase one should
have a Goldstone mode with one
 velocity  related to the spin rotation in the plane of the spiral, and two
Goldstone modes with another velocity related to the rotations of the plane of
the spiral around $X$ and $Y$ axes ~\cite{zuber}. The former Goldstone mode
results in the
divergence of the total static susceptibility $\chi^{+-} (q)$ at the wave
vector
 $q = -\bar Q$,
while the latter lead to divergences of the total $\chi^{zz} (q)$
at the wave vectors ${\bf \pm \bar Q}$.

The spectrum of collective excitations is determined by the poles of
dynamic susceptibility
\begin{equation}
\chi^{ij}_{q} = i \int dt e^{i\omega t}
\langle TS^i_{q}(t)S^j_{-q}(0) \rangle
\end{equation}
Here $S^i$ represents either one of the three spin densities, $S^+, S^-,
 S^Z$, or the charge density $\rho$. For the $(\pi,\pi)$ state, fluctuations
in transverse spin channels are completely decoupled from fluctuations in
the density and longitudinal spin channels. Dynamical susceptibility is then
a $2\times2$ problem. For planar spiral states, however,
all four channels are coupled, and
the dynamical
susceptibility has to be found by solving a set of four coupled
Dyson equations. Doing the standard SDW manipulations, we obtain that the
poles of dynamical susceptibility are given by solving $D(q,\omega)
 =0$ where $D (q,\omega)$ is the determinant of a $4\times4$ matrix given by
\[
\left(\begin{array}{clcr}
1 - U \chi^{+-}_{q,-q} & -U \chi^{--}_{q+2{\bar Q},-q} & - U \sqrt{2}
 \chi^{z-}_{q +{\bar Q},-q} & U \sqrt{2}\chi^{\rho -}_{q+{\bar Q},-q} \\
-U \chi^{++}_{q, -(q+2{\bar Q})} & 1 - U \chi^{-+}_{q+2{\bar Q},-(q+2{\bar Q})}
&
-U \sqrt{2}\chi^{z+}_{q+{\bar Q},-(q+2{\bar Q})} & U \sqrt{2}\chi^{\rho
+}_{q+{\bar Q},-(q+2{\bar Q})} \\
-U \sqrt{2} \chi^{+z}_{q,-(q+{\bar Q})} & -U \sqrt{2}
\chi^{-z}_{q+2{\bar Q},-(q+{\bar Q})} & 1 -
 2U \chi^{zz}_{q+{\bar Q},-(q+{\bar Q})} & 2U \chi^{\rho z}_{q+{\bar
Q},-(q+{\bar Q})} \\
-U \sqrt{2} \chi^{+ \rho}_{q,-(q+{\bar Q})} & - U \sqrt{2}
\chi^{- \rho}_{(q+2{\bar Q}),-(q+{\bar Q})} & -
2U \chi^{z\rho}_{q+{\bar Q},-(q+{\bar Q})} &
1 + 2U \chi^{\rho \rho}_{q+{\bar Q},-(q+{\bar Q})}\\
\end{array}
\right)
\]
The expressions for the irreducible
 susceptibilities are presented in the appendix A. A similar expression for
$D(q,\omega)$ was recently obtained by Cote and Tremblay~\cite{tremblay}
 in their SDW analysis
of collective excitations in 2D Hubbard model on a triangular lattice at
half-filling
For our present consideration, it is essential that at zero frequency,
$\chi^{+z}_{q,-(q+{\bar Q})} = \chi^{-z}_{q+2{\bar Q},-(q+{\bar Q})}
 = \chi^{z\rho}_{q+{\bar Q},-(q+{\bar Q})} =0$ {\em at any $q$}, and therefore
static
transverse spin fluctuations {\it decouple} from the longitudinal spin
fluctuations and density fluctuations (we remind that spin ordering  is in
the XY plane). We now consider the transverse and longitudinal fluctuations
separately.

\subsection{Longitudinal spin fluctuations}

 Consider first the solution for the
density and longitudinal spin fluctuations.
 At $q=-\bar Q$, the evaluation of the expressions in the appendix A yields
at zero frequency
 $\chi^{+\rho}_{-{\bar
Q},0} = \chi^{-\rho}_{{\bar Q},0} = \chi^{\rho,\rho}_{-{\bar Q},{\bar Q}} =
z/8U$,
{}~$\chi^{++}_{-{\bar Q},-{\bar Q}} =
\chi^{--}_{{\bar Q},{\bar Q}} = (z-4)/8U$, and
$\chi^{+-}_{-{\bar Q},{\bar Q}} - \chi^{--}_{{\bar Q},{\bar Q}} = \sum_k
1/(2E^{-}_k) \equiv 1/U$. Elementary manipulations then show
 that there is indeed a Goldstone mode at
$q=-{\bar Q}$. Expanding around this point, we obtain
after straightforward
but lengthy calculations, that to quadratic order in $q+{\bar Q}$
\begin{equation}
D (q,\omega) = \frac{2 t^2}{U^2}~ (q + {\bar Q})^2 \left(1 -
\frac{z_{\omega}}{2}\right) -
\frac{\omega^2}{U^2}.
\end{equation}
The $q+ {\bar Q}$ term comes from the expansion of
$1 - U\chi^{+-}_{q,-q} + U \chi^{++}_{q,-(q+2{\bar Q})}$, and
$\omega^2$ term comes from $\chi^{+z}$ and $\chi^{-z}$.
We also introduced
$z_{\omega} = 4U \chi_{2D} (\omega)$ where $\chi_{2D} (\omega)$ is the
susceptibility of a 2D Fermi gas {\it {at finite frequency}}
\begin{equation}
\chi_{2D} (\omega) = \frac{\sqrt{m_{\parallel} m_{\perp}}}{2\pi} \left[ 1 -
\left(\frac{\delta^2}{\delta^2 - 1} \right)^{1/2}\right]
\end{equation}
where to the leading order in the hole density,
\begin{equation}
\delta = \frac{\omega^2 m_{\parallel} m_{\perp}}{(q + {\bar Q})^2 p^2_{F}} .
\end{equation}
and $p_F \sim \sqrt{x}$ is
the Fermi momentum of holes.
At zero frequency we indeed have $z_{\omega} = z$.

The explicit expression for the
total longitudinal susceptibility, ${\overline\chi}$, is
\begin{equation}
{\overline \chi}^{+-} (q,\omega) = \frac{A (q,\omega)}{D (q,\omega)}
\end{equation}
The numerator can be evaluated right at $q= - {\bar Q}, \omega =0$
where it reduces to
\begin{equation}
A (-{\bar Q},0) =
 (\chi^{+-}_{-{\bar Q},{\bar Q}} -  \chi^{++}_{-{\bar Q},-{\bar Q}})
{}~\left[(1 - U\chi^{+-}_{-{\bar Q},{\bar Q}})( 1 +
2U \chi^{\rho,\rho}_{-{\bar Q},{\bar Q}}) + 2U^2
\chi^{+\rho}_{-{\bar Q},0} \chi^{-\rho}_{{\bar Q},0}\right]
\end{equation}
Substituting the values for irreducible susceptibilities, we obtain that
$A= 1/2U$ and does not depend on $z$. For the total longitudinal
susceptibility we then have
\begin{equation}
({\overline \chi}^{+-}_{q,-q})^{-1} =
J (q + {\bar Q})^2 \left(1 - \frac{z_{\omega}}{2}\right) - \frac{\omega^2}{2J}
\label{chistat}
\end{equation}
 This expression is valid for $(\pi,Q)$
phase and for $(Q,Q)$ phase at $q_x = q_y$. We see that the stiffness for
 longitudinal fluctuations is
$\rho_{L} \propto (1-z/2)$. For the $(\pi,Q)$ phase ($1<z<2$),
 the stiffness is positive,
while for the $(Q,Q)$ phase ($z>2$) it is {\it {negative}}. This last result
implies that the homogeneous $(Q,Q)$ phase is in fact unstable. This agrees
with our observation in the previous section that $\partial E^{(Q,Q)}/\partial
x^2$ is negative.

The negative longitudinal stiffness of the $(Q,Q)$ phase was earlier obtained
by Dombre~\cite{dombre} in the macroscopic calculations in the framework of the
Shraiman-Siggia model. He argued that this
instability leads to a formation of domain walls, but can be prevented by
a long-range Coulomb interaction.  Phase separation at large $z$ is also a
possibility~\cite{Rice}. We however have not studied inhomogeneous spin
configurations.

It is also essential to observe that $z_{\omega}$
 behaves as a constant ($=z$) only at frequencies
comparable to $(q + {\bar Q}) p_F$;
 at larger frequencies, dynamical susceptibility
$\chi_{2D} (\omega)$ rapidly decreases, and near a spin-wave pole, $\omega^2
 \sim 2J^2 (q + {\bar Q})^2$,
 we have $\chi_{2D} (\omega) \sim p_F |q + {\bar Q}|/\omega \ll 1$.
Then, near the pole, to the lowest order in the hole density we have
\begin{equation}
{\overline \chi}^{+-} \sim (c^2 (q + {\bar Q})^2 - \omega^2)^{-1}~,
\label{chizzz}
\end{equation}
where the spin-wave velocity, $c=\sqrt{2}J$, is the same as in the
 $(\pi,\pi)$. This result agrees with the macroscopic consideration by
Shraiman and Siggia~\cite{boris} and with the Schwinger-boson analysis by Gan,
Andrei and Coleman~\cite{piers}.

\subsection{Transverse spin fluctuations}

We now  consider the magnetic susceptibility $\chi^{zz}_{q,q}$
associated with the
out-of-plane fluctuations. We found above that this channel is
 coupled to density and longitudinal spin fluctuations only dynamically.
For the full static susceptibility  we then have a simple  RPA formula
\begin{equation}
{\overline\chi}_{q,-q}^{zz} = \frac{\chi^{zz}_{q,-q}}{1
-U\chi^{zz}_{q,-q}}
\end{equation}
{}From the above considerations, we expect the
 Goldstone modes in
 $\overline{\chi}_{q,-q}^{zz}$ to be at ${q=\pm \bar Q}$.
Consider first the $(\pi,Q)$ spiral.
Using eq. (\ref{chizz}) from the appendix A,
and expanding near ${q=\pm \bar Q}$,
to the lowest order in the hole density we obtain
\begin{equation}
\chi_{q,-q}^{zz} = \frac{1}{2U} + \frac{x^2}{U}~\left(\frac{t {\bar q}}{U
x}\right)^2 \left( 1 - \frac{{\tilde q}^2}{{\bar q}^2}\right) -
\frac{x^2}{U}~\left(\frac{t {\bar q}}{U
x}\right) \left( 1 - \frac{{\tilde q}^2}{{\bar q}^2}\right) ,
\label{aaa}
\end{equation}
where ${\tilde q} = \pi - q$ and ${\bar q} = \pi - Q$.
The second term comes from the integration over the regions away from
pockets, while the last term comes from the integration inside the hole
pockets.
We see that at ${\tilde q} = \pm {\bar q}$, i.e., at $q = \pm Q$,~ $\chi^{zz}$
precisely equals $1/2U$, and ${\overline\chi}^{zz}$ diverges as it indeed
should.  At the same time, neither of the last two terms in (\ref{aaa})
 has a form of a quadratic expansion
{\it {around}} $q = \pm Q$. To the lowest order in hole density,
$ {\bar q} = U x/t$, and the last two terms in (\ref{aaa}) cancel each other
at any $q \sim Q$. This means that ${\overline\chi}^{zz}$ is infinite
in a whole range of momenta which in turn implies that there exists an infinite
number of other states which are degenerate in energy with the spiral states to
the leading order in hole density. In appendix B we show that this degeneracy
is
in fact a quite general phenomenon and it exists for an arbitrary ratio of
$t,t^{\prime}$ and $U$.  Furthermore, we found that at least to the
leading order
in $t/U$, $\chi_{q,-q}^{zz} \equiv 1/2U$ for $q = (\pi, {\tilde q}_y)$
and {\it arbitrary} ${\tilde q}_y$. This last result means that there exist
a whole line of zero modes in ${\overline\chi}^{zz}$. At $q =
({\tilde q}_x, {\tilde q}_y)$, we found that $\chi_{q,-q}^{zz} = 1/2U +
 O(({\tilde q}_x)^4)$. For
typical ${\tilde q}_x \sim {\bar q} \sim x$, the last term is $O(x^4)$, i.e.,
it contains two extra powers of $x$ compared to the terms we consider.

We will discuss the set of degenerate states in the next
section, and here merely notice that the degeneracy is not
related to any kind of broken symmetry and, therefore, should be lifted by
higher-order terms in the expansion in the hole density which we are proceeding
to discuss.

It is not difficult to make sure that the contributions to $\chi^{zz}$ from the
regions far from the hole pockets form regular series in powers of ${\tilde
q}^2$; odd powers of ${\tilde q}$ disappear due to momentum integration.
As typical ${\tilde q} \sim {\bar q}$, the next-to-leading order
terms have an extra factor of ${\bar q}^2 \propto x^2$. On the other hand, the
expansion near pockets involve only fermions with momenta
near $k =(\pi/2,\pi/2)$
and $k = (-\pi/2,\pi/2)$, i.e.,
there is no summation over
momenta.
 As a result, the next subleading term
in $\chi^{zz}$ from hole pockets has an extra power of $x$
rather than $x^2$. We computed this
term explicitly by expanding in (\ref{chizz})
and in the
ground state energy (from which we extract ${\bar q}$)
 beyond the leading order in hole
density, and obtained
\begin{equation}
\bar q = \frac{Ux}{t} \left( 1 - \frac{\pi}{8}~x~\frac{m_{\parallel} +
m_{\perp}}{\sqrt{m_{\parallel} m_{\perp}}}\right)
\label{eeee}
\end{equation}
and
\begin{equation}
\chi_{q,-q}^{zz} = \frac{1}{2U} - \frac{2 x^3}{U}
\left( 1 - \frac{{\tilde q}^2}{{\bar q}^2}\right)~ \left[\left(1 -
\frac{2}{z}\right) - \frac{{\tilde q}^2}{{\bar q}^2}\right]
\label{chiinst}
\end{equation}
Notice that the correction
term in ${\bar q}$ which explicitly depends on the
mass ratio does not show up in the expression for $\chi^{zz}$.

We see that the Goldstone mode in ${\overline \chi}^{zz} \sim (1 - 2U
\chi^{zz})^{-1}$ at ${\tilde q} = \pm {\bar q}$ survives as it
should, but $({\overline \chi}^{zz})^{-1}$ still {\it {does not}}
 have a form of a quadratic expansion {\it around} the Goldstone points
(Fig.~\ref{fig_chiinst}).
For $1<z<2$, when the $(\pi,Q)$ phase is a candidate for the ground state,
$1 - 2U \chi^{zz}$, and hence, $\overline{\chi}^{zz}$, is negative
in the region ${\tilde q}^2 < {\bar q}^2$. This implies that the $(\pi,Q)$
planar spiral phase is also unstable, and one should look for other
candidates for the ground state.

For completeness,  consider also
the transverse susceptibility in the $(Q,Q)$ phase. At ${\tilde q}_x \neq
 {\tilde q}_y$,  the total transverse susceptibility is positive
already at the quadratic order in ${\tilde q}$:
\begin{equation}
1-2U\chi^{zz} = \frac{t^2}{U^2}({\tilde q}_x-{\tilde q}_y)^2.
\end{equation}
However, along the
Brillouin zone diagonal, at ${\tilde q}_x = {\tilde q}_y ={\tilde q}$,
$(\overline \chi)^{-1}$  is
again zero,  to order $O(x^2)$, at arbitrary ${\tilde q}$.
We performed calculations
to order $O(x^3)$ and obtained
\begin{equation}
1-2U\chi^{zz} = 4x^3\left(1-\frac{{\tilde q}^2}{{\bar q}^2}\right)
\left[1-\frac{2}{z} - \frac{4{\tilde q}^2}{{\bar q}^2}\right]
\end{equation}
The total transverse susceptibility ${\overline\chi}^{zz} = \chi^{zz}/(1 - 2U
\chi^{zz})$ then
has one zero associated with the Goldstone mode at
${\tilde q} = \bar q$, and another `` accidental'' zero at
\begin{equation}
\tilde q = \bar q \sqrt{\frac{1}{4}\left(1-\frac{2}{z}\right)}.
\end{equation}
Between these two zeros the total
static transverse susceptibility is negative,
indicating an instability. Contrary to the previously found instability  of the
$(Q,Q)$ state towards domain-wall formation, the latter instability
 is unlikely to be removed by
including the long-range component of Coulomb interaction.  We however have not
performed any further calculations for the $(Q,Q)$ phase.

The result that the planar spiral phase is unstable contradicts  the
assumption made by Shraiman and Siggia that the subleading terms in the
expansion over $x$ stabilize the planar phase. On the contrary, our results
indicate that they do not.

\subsection{Uniform susceptibility}

We conclude this section with a brief consideration of the uniform
susceptibility of the spiral states. The key point here is that the
spiral ordering in the $XY$ plane
away from half-filling couples the fluctuations along $X$ and $Y$ directions,
which at half-filling constituted longitudinal and transverse fluctuations.
We have found in Sec.~\ref{m-f}, that in the $(\pi,\pi)$ state, longitudinal
fluctuations acquire a correction proportional to the Pauli susceptibility,
while transverse
fluctuations do not. In a spiral phase, these fluctuations are coupled, and
as a result, the uniform static susceptibility of the ordered state
acquires a finite correction immediately away from half-filling
\begin{equation}
{\overline\chi}^{+-} (0,0)  = \frac{1}{8J} + \frac{{\tilde\chi}_{2D}}{4} ,
\end{equation}
where
\begin{equation}
\tilde\chi_{2D} = \lim_{q \rightarrow 0}
\frac{1}{N} ~{\sum_{\stackrel{E^{d}_{k+q} <\mu}{E^{d}_{k-{\bar Q}} >\mu}}}
 \frac{1}{E^{d}_{k-{\bar Q}} - E^{d}_{k+q}}
\label{bbb}
\end{equation}
For ${\bar Q}= (\pi,\pi)$, $\tilde\chi_{2D} = \chi_{2D} = \sqrt{m_{\parallel}
m_{\perp}}/2\pi$. However, for finite $\bar q = Ux/t$,
 the $t{\bar q}$ term in the denominator in (\ref{bbb}) is the dominant one,
and
performing calculations we obtain
\begin{equation}
\tilde\chi_{2D} = \frac{1}{8J}~ \frac{2J^2}{t^2}
\end{equation}
We see, therefore, that the step-like correction to the uniform susceptibility
is relatively small at $J \ll t$. Notice also that fluctuations in the $Z$
direction are decoupled from the $XY$ fluctuations, and hence
${\overline\chi}^{zz}_{00} = (1/8J) + O(x)$ without any step-like corrections.

We now proceed to the analysis of the states degenerate with the
planar spiral to the lowest order in the hole density.

\section{Non-coplanar states}

To specify the set of degenerate states,
we first return to our results obtained to order $O(x^2)$ and
observe that the zero modes in $\chi^{zz}$
are centered around $(\pi,\pi)$. A zero mode in the transverse susceptibility
at
$(\pi,\pi)$ implies that the system is indifferent towards generation
of a spontaneous commensurate antiferromagnetic order along
$Z$ direction in addition to the incommensurate spin ordering in the XY plane
(Fig.~\ref{nonplanara}).
We, therefore, consider a set of states having
two different SDW order parameters $\Delta_{\perp} =
 U\langle S_\perp\rangle $, and
$\Delta_{\parallel} = U\langle S_\parallel\rangle $, where
 $\langle S_\perp \rangle $ and
$\langle S_\parallel \rangle $ are the magnitudes of the off-plane and in-plane
components of the on-site magnetization, respectively.
Performing the mean-field decoupling of the interaction term
and the diagonalization of the Hubbard
Hamiltonian, we obtain
\begin{equation}
H=\sum_k \overline E_k(c^{\dag}_k c_k - d^{\dag}_k d_k)
\end{equation}
with the energy in the valence (d) and conduction (c) bands given
by
\begin{equation}
\overline E_k=\sqrt{(E_+ + \sqrt{\Delta_\parallel^2+E_-^2})^2
+\Delta_\perp^2}
\label{new_sp}
\end{equation}
The ground state energy, to order $O(x^2)$, is given by
\begin{equation}
E_{\Delta_\perp} = \frac{U x^2}{z} - t q^* x + \frac{(q^*)^2 t^2}{2U}
\label{ee}
\end{equation}
where  $q^{*} = \bar q(\Delta_\parallel/\Delta)$, and
$\Delta^2 = \Delta_\perp^2 + \Delta_\parallel^2$. For $U \gg t$ we indeed have
$\Delta \approx U/2$. We also assumed in (\ref{ee}) that
$\Delta_{\parallel} \gg x^{1/2}$.

Since we now have two order parameters, there are also two
self-consistency conditions. The condition on the out-of-plane
order parameter $\Delta_\perp$ is:
\begin{equation}
\frac{1}{U} = \sum_k \frac{1}{2\overline E_k},
\end{equation}
and the condition on the in-plane order parameter $\Delta_\parallel$ is:
\begin{equation}
\frac{1}{U} = \sum_k \frac{1}{2\overline E_k}
\left(1+\frac{E_+}{\sqrt{\Delta_\parallel^2+E_-^2}}\right) ,
\label{self_cons}
\end{equation}
In the limit $\Delta_\perp \rightarrow 0$, the latter expression reduces
to (\ref{self-conf}) as it should. However, for any nonzero $\Delta_\parallel$,
the two self-consistency conditions have to be satisfied simultaneously,
 which implies that the inverse pitch of the spiral
is no longer a free parameter.
Specifically, the
compatibility of the two conditions requires that
\begin{equation}
\sum_k \frac{E_+}{\sqrt{\Delta_\parallel^2+E_-^2}E_k} = 0
\label{self_cons2}
\end{equation}
However, solving this equation to order $O(x^2)$, we find that
\begin{equation}
q^*= \frac{U}{t}x
\end{equation}
which is exactly what one would obtain by simply minimizing the
ground state energy (\ref{ee}) with respect to $q^{*}$.
As a result, substituting $q^*$ into (\ref{ee}), we obtain that the
ground state energy {\it does not} depend on $\Delta_{\perp}$
\begin{equation}
E_{\Delta_\perp} = U x^2 ~\left(\frac{2}{z} -1 \right)~\equiv E^{(\pi,Q)}
\end{equation}
Clearly then, all states with finite $\Delta_{\perp}$ are degenerate in energy
with the planar spiral, and we have to go beyond the leading order in
$x$ to see which state actually has the lowest energy.

The calculations to order $O(x^3)$ proceed in the same way as in the previous
section. We skip the details and focus on the results. For the
 inverse pitch of the spiral we obtained from the consistency condition
(\ref{self_cons2})
\begin{equation}
q^* = \frac{U}{t}~x \left[1+2x\left(
1-\frac{\Delta^2+\Delta_{\parallel}^2}{2\Delta_{\parallel}^2}
\frac{2}{z}\right)\right]~,
\label{exprq}
\end{equation}
Here we again assumed that
$\Delta_{\parallel}/\Delta \gg x^{1/2}$.
We see now that the values of ${\bar q}$ at $\Delta_{\perp} =0$ (eq.
(\ref{eeee}) and at
 $\Delta_\perp \to 0$ (eq. (\ref{exprq})) are different to order $O(x^3)$.
 The reason is, of
course, that in the case of a planar spiral there is only one self-consistency
condition to satisfy.
The energy of the noncoplanar phase, at $\Delta_\parallel/\Delta \gg x^{1/2}$,
is given by
\begin{equation}
E_{\Delta_\perp} = E^{(\pi,Q)} + \frac{t}{2U}\left[q^* - \frac{Ux}{t}\right]^2
+ Ux^3\left(\frac{\Delta_\perp}{\Delta_\parallel}\right)^2\left(1-\frac{2}{z}
\right)
\label{en1}
\end{equation}
where $q^*$ is given by (\ref{exprq}).
Let us first discuss the
 second term in the r.h.s. of (\ref{en1}).
This is a positive contribution to the energy related  to the fact that
$q^*$ is no longer a free parameter. At $\Delta_{\perp}
\rightarrow 0$, the last term in the r.h.s. of (\ref{en1}) disappears.
and the energy of the noncoplanar state turns out to be larger than that
of the $(\pi,Q)$ state. Clearly then,
{\it very}
 close to $\Delta_{\perp} =0$, a simple noncoplanar state that
we consider is not the best choice. This in fact is
 consistent with the form of the susceptibility
in the spiral phase which has a minimum at some momentum different from
$(\pi,\pi)$ (see Eq.(\ref{chiinst})).
 However, we also see that this energy
difference is $O(x^4)$, and for all $\Delta_{\perp} \geq x^{1/2}$, the second
term in the  r.h.s.
of (\ref{en1}) can be neglected compared to the third term
which is of the order $x^3$ and negative for  $z<2$ which we consider.
This latter term increases with $\Delta_{\perp}$. Moreover, eq. (\ref{en1})
does
not contain a
``restoring force'' (i.e., terms $\sim x^3
\Delta^{4}_{\perp}$). Therefore, by making $\Delta_{\perp}$ larger and larger,
we can continuously decrease the ground state energy as long as eq. (\ref{en1})
remains valid, i.e., as long as
$\Delta_\parallel/\Delta \gg x^{1/2}$. When $\Delta_{\perp}$ nearly reaches
$\Delta$, and
$\Delta_{\parallel}/\Delta$ becomes comparable with $x^{1/2}$,
the expression for the ground state
 energy of the noncoplanar state becomes more complex. In this situation, we
found
\begin{equation}
E_{\Delta_\perp} = \frac{U x^2}{z} +
 \frac{Ux^2}{2}\left[(\alpha\beta)^2 + 2\beta^2
- \frac{\beta z}{6}f(\alpha)\right],
\end{equation}
where
\begin{eqnarray}
\alpha = \frac{\Delta_\parallel}{\Delta_\perp x^{1/2}},~~\beta =
\frac{q^* t}{Ux}\frac{\Delta_\perp x^{1/2}}{\Delta_\parallel}\nonumber\\
f(\alpha) = \left(\alpha^2 + \frac{8}{z}\right)^{3/2} - \alpha^3
\label{en2}
\end{eqnarray}
For $\Delta_\parallel/\Delta \gg x^{1/2}$, $\alpha \gg 1$, and expanding in
$1/\alpha$ in (\ref{en2}) we recover (\ref{en1}).

The solution of the self-consistency condition (\ref{self_cons2}) is also more
complex for $\Delta_\parallel/\Delta \sim  x^{1/2}$. We found
\begin{equation}
\beta = \frac{z}{2}\left[\sqrt{\alpha^2+\frac{4}{z}} - \alpha\right].
\end{equation}
At $\alpha \gg 1$ (but still, $\Delta_{\parallel} \ll \Delta$),
 this expression reduces to (\ref{exprq})

Substituting $\beta$ into (\ref{en2}), we obtain the ground state
energy as a function of a single free parameter $\alpha$.
 The equilibrium value of
$\alpha$ (and, hence, of  $\Delta_{\perp}$) can now be obtained by a simple
minimization of the energy.
In a general case, the solution of $ dE_{\Delta_\perp}/d\alpha =0$
is rather involved, but
for $z$ close to $2$, a  simple analytical solution is possible.
The point is that at $z \approx 2$, the equilibrium
value of $\alpha$ is large ($\sim (2 -z)^{-1/2}$), so we can expand (\ref{en2})
in powers of $1/\alpha^2$. We then obtain
\begin{equation}
E_{\Delta_\perp} = E^{(\pi,Q)} + Ux^2\left[\frac{1}{\alpha^2}
\left(1-\frac{2}{z}
\right) + \frac{7}{24\alpha^4}\right],
\label{en3}
\end{equation}
The energy has a minimum at
\begin{equation}
\alpha = \sqrt{\frac{7z}{12(2-z)}}.
\end{equation}
For the equilibrium noncoplanar state we thus have
\begin{eqnarray}
E_{\Delta_\perp} =
\frac{Ux^2}{2z}(2-z)\left[1-\frac{6}{7}(2-z)\right],
\nonumber\\
\bar q = \sqrt{\frac{12}{7}~x\left(\frac{2}{z}-1\right)},~
\Delta_\parallel = \Delta\sqrt{\frac{7}{12}~\frac{xz}{2-z}}.
\end{eqnarray}
We see that the energy of the noncoplanar state is substantially
lower than that of the
$(\pi,Q)$ state. What is more,
the energy gain in the equilibrium state scales as $x^2$, instead of $x^3$ as
in (\ref{en1}). This implies that
the equilibrium state with
$\Delta_\parallel \sim \sqrt{x}$, strictly speaking, does not belong
 to the original set of degenerate spin configurations, and
could be selected already in the calculations to order $O(x^2)$. The discovery
of the degenerate set of states and of the instability of the planar spiral
gave us, nevertheless, a hint where to look for the minimum of the
energy. Notice also that $\partial E_{\Delta_\perp}/\partial x^2 >0$, i.e.,
there is no instability towards phase separation.

Another important point concerns the chirality of the novel state. Although
the spin configuration in this state is noncoplanar, it is not chiral in the
sense that there is no flux through a plaquette~\cite{wwz}. In other words,
although, the triple product of three adjacent spins along the $Y$ direction,
$\vec S_{i\,j-1}\cdot(\vec S_{i\,j}\times \vec S_{i\,j+1}) \neq 0$, the
triple product of spins lying in the vertices of a minimal triangle is always
zero, since all spins along the rows in the $X$ direction are parallel to each
other. This fact distinguishes our noncoplanar state
from the double spiral considered in~\cite{lee}b, which has staggered
chirality. We emphasize however that, at least in the SDW approximation, our
noncoplanar state has $lower$ energy than the double spiral.

\section{Magnetic susceptibility of the equilibrium state}

In the presence of a commensurate antiferromagnetism along $Z$-axis, the
equation for magnetic instability becomes  more complex as now bare
susceptibilities with the momentum transfer $(\pi,\pi)$ are also finite,
and the total susceptibility becomes $8 \times 8$ problem. In view
of this, we only computed the susceptibility to the leading order in
$\Delta_{\perp}$. We found that the compatibility condition (\ref{self_cons2})
of the two self-consistent equations at $\Delta_{\perp} \rightarrow 0$
 is equivalent (to order $O(x^3)$)
to the condition that ${\overline\chi}^{zz}$ diverges at
$(\pi,\pi)$. (see Fig~\ref{fig_chi_zz}). This extra zero mode exists because
at $\Delta_{\perp} = 0$, the ground state energy as a function of
$\Delta_{\perp}$ has an extremum (maximum).
We have not performed calculations at
$\Delta_{\perp} \geq \Delta_{\parallel}$, but we believe it
 plausible that the spin susceptibility evolve with $\Delta_{\perp}$ as it is
shown in Fig~\ref{fig_chi_zz}. The equilibrium state is an energy minimum (at
least, local), and we expect that the static susceptibility of this state is
positive, diverging only at the Goldstone points.  There exists,
 however, a subtlety in determining the locations of zero modes in this
configuration, so we will explicitly follow the
recipe of the Goldstone theorem~\cite{zuber}. This theorem states that
if $\hat J$ is a generator of a
symmetry transformation, and
the commutator $[\hat A,\hat J]$, where $A$ is some operator,
 has a non-zero
average value in the ground state, then
 the correlator $\langle TA^{\dag}A \rangle$ diverges at $\omega = 0$. The
residue of the quasiparticle pole near the Goldstone point is
proportional to $[\hat A,\hat J]^2$.
In our case the corresponding operators and correlators are the following:
\begin{enumerate}
\item
Rotation about the $X$-axis: $\hat J = S_0^x$. If $\hat A = S_{\pi}^y$
{}~($\pi \equiv (\pi,\pi)$)),
 then $[\hat A,\hat J] = -iS_{\pi}^z \sim \Delta_\perp$, and therefore
$\chi^{yy}_{\rm st}(\bf q) $ diverges
at ${\bf q = \pi}$. The residue of
the pole is proportional to  $\Delta_\perp^2$.
 If instead
we choose $\hat A = S_{\pm \bar Q}^z$,
 then $[\hat A, \hat J] = iS_{\pm \bar Q}^y$, and we find a divergence
 in $\chi^{zz}_{\rm st}(\bf q)$ at ${\bf q = \pm \bar Q}$;
the residue of the pole is proportional to $\Delta_\parallel^2$.
\item
Rotation about the $Y$-axis: $\hat J = S_0^y$.
 This case is analogous
to the previous one. The divergences occur in
$\chi^{xx}_{\rm st}( \pi)$ with the residue of the pole
$\propto \Delta_\perp^2$,
and in $\chi^{zz}_{\rm st}({\bf \pm \bar Q})$ with the residue
$\propto \Delta_\parallel^2$.
\item
Rotation about the $Z$-axis: $\hat J = S_0^z$. Choosing
$\hat A = S_{\bar Q}^+$ and $\hat A = S_{-\bar Q}^-$ we find
divergences in $\chi^{+-}_{\rm st}({\bf \bar Q})$
and $\chi^{-+}_{\rm st}({- \bf \bar Q})$, in both cases the residue of the pole
is proportional to $\Delta_\parallel^2$.
\end{enumerate}
Combining these results, we find that in  the equilibrium noncoplanar state,
the in-plane static susceptibility $\chi^{+-}$  in fact
has two poles, one at at $q = (\pi,\pi)$ and the other at $q = -\bar Q$.
Then, for $q$ not too far from $(\pi,\pi)$, this susceptibility can be
approximated as
\begin{equation}
\chi^{+-}({\bf q}) \approx \frac{\chi_\pi}{({\bf q}-{ \pi})^2}+
 \frac{\chi_{\bar Q}}{({\bf q + \bar Q})^2}~,
\end{equation}
where the residues $\chi_\pi$ and $\chi_{\bar Q}$ are proportional to
$\Delta_\perp^2$ and $\Delta_\parallel^2$ respectively. Because
$\Delta_\parallel \sim \sqrt{x}$, the
residue of the pole at the
incommensurate wave vector $\bf q=-{\bar {\bf Q}}$ is
 suppressed with respect to the pole at the commensurate wave
vector $(\pi,\pi)$, and the form of the static
susceptibility is very similar to that
for the $(\pi,\pi)$ state (Fig.~\ref{fig_chi_pm}).

 The out-of-plane static susceptibility $\chi^{zz}$ also has
two poles
\begin{equation}
\chi^{zz}({\bf q}) \approx \frac{\chi_z}{({\bf q - \bar Q})^2}+
 \frac{\chi_z}{({\bf q + \bar Q})^2}~,
\end{equation}
but here both poles are suppressed as the residue,
$\chi_z$, is proportional to $\Delta_\parallel^2$, and, therefore, to $x$.
Again, the form of the susceptibility is very similar to that for the
$(\pi,\pi)$ state. On the contrary, in the planar spiral state,
both transverse and longitudinal susceptibilities have Goldstone modes
at $q = \pm {\bar Q}$ with the residue of the pole proportional to the total
on-site magnetization (Fig.~\ref{fig_chi_zz}).

\section{Conclusions}
\label{concl}

Here we summarize the main results of this paper. We used spin-density-wave
formalism and  studied
various magnetic phases of the 2D Hubbard model
at low doping when a long-range magnetic order is still present.
  We found that the equilibrium spin configuration
 depends on the value of the dimensionless
parameter $z = 4T \chi^{2D}$, where $T$ is the scattering amplitude of two
holes
($T = U$ in the mean-field approximation), and $\chi^{2D}$ is the Pauli
susceptibility of holes which at low doping occupy pockets located at
$(\pi/2,\pi/2)$ and symmetry related points in the Brillouin zone. In $2D$,
Pauli susceptibility does not depend on the carrier concentration:
$\chi^{2D} = \sqrt{m_\parallel m_\perp}/2\pi$. We found
that for $z <1$, the commensurate antiferromagnetic state is stable.
For $z >2$, the spiral $(Q,Q)$ phase is the equilibrium configuration at the
mean-field level, but we found that this configuration has negative
longitudinal stiffness and therefore is unstable against domain wall
formation. This result agrees with the macroscopic analysis in~\cite{dombre}.
The intermediate case, $1<z<2$, is the most interesting
from a theoretical point of view. Here
we found that the equilibrium configuration at the mean-field level
is a $(\pi,Q)$ spiral introduced by Shraiman and Siggia.
 The longitudinal stiffness in this configuration is
positive,  but the transverse stiffness
vanishes to the leading order in hole density. This in turn implies that
the spiral phase is degenerate in energy with many other spin configurations,
and the equilibrium state only appears as an ``order from disorder'' effect.
We performed calculations beyond the leading order in the hole density and
found
that the equilibrium state is not a planar spiral but rather a noncoplanar
spin configuration which contains both, $(\pi,\pi)$ antiferromagnetism along
one direction in spin space, and $(\pi,Q)$ spiral in the orthogonal plane.
The latter result suggests a novel
scenario of spin reorientation with doping for $1<z<2$, different from the one
suggested by Shraiman and Siggia. In their
picture, upon doping
spins remain in the same plane as at half-filling, but
twist into a spiral with incommensurate momentum $(\pi,Q)$.
In our scenario, the commensurate antiferromagnetic ordering (same as at
half-filling) does not vanish, as doping only
introduces a
transverse component of the order parameter which forms a spiral in the
plane perpendicular to the direction of the commensurate order. This transverse
component  is small to the extent of $x$, and the low-$T$ behavior
at finite doping remains nearly the same as in the commensurate
antiferromagnet~\cite{commensur,cs}(Fig.~\ref{nonplanarb}).

It is essential that
our analysis has been performed only
for frequencies smaller than
the energy scale $\Delta E$ associated with the lifting of the degeneracy.
At larger  frequencies,
 the static selection may be irrelevant, and
one has to solve the full dynamical problem which presents a technical
challenge.

The above analysis is valid for the magnetically ordered phase.
We therefore cannot pretend to resolve the known discrepancy
between neutron scattering and NMR experiments
in $La_{2-x}Sr_{x}CuO_{4}$~\cite{nmr,incommen},
both of which have been performed well
inside the metallic phase. We merely note that,
as neutron data indicate, the incommensurability at $(\pi,Q)$ observed
at 7.5\% and 14\% doping {\it is not} correlated with the magnetic behavior
in the ordered phase. This implies that our result that in the ordered state,
susceptibility is always peaked at $(\pi,\pi)$, does not contradict the
neutron data.

It is our pleasure to thank  A. Abanov, V. Barzykin, P. Chandra, P. Coleman,
R. Gooding, R. Joynt, D. Khomskii, D. Pines, S. Sachdev, Q. Si and
A. Sokol for useful
discussions. The work was supported by the University of Wisconsin-Madison
Graduate School and Electric Power Research Institute.

\section{appendix A}

In this appendix we present the results for the irreducible susceptibilities in
the spiral phases. Each of the susceptibilities below was obtained by the
standard SDW manipulations.
\begin{eqnarray}
\chi^{zz,\rho\rho}_{q,-q} &=&
 \frac{1}{8N} \sum_{E^{d}_k <\mu}~ \left[1 -
\frac{\epsilon^{-}_k \epsilon^{-}_{k+q} \mp \Delta^2}{E^{-}_k
E^{-}_{k+q}}\right] \left(\frac{1}{E^{c}_{k+q} -
 E^{d}_k - \omega} +
\frac{1}{ E^{c}_{k+q} - E^{d}_k + \omega}\right) +
 \nonumber \\
&&  \frac{1}{8N} \sum_{\stackrel{E^{d}_{k+q} <\mu}{E^{d}_{k} >\mu}}
 \left[1 +
\frac{\epsilon^{-}_k \epsilon^{-}_{k+q} \mp \Delta^2}{E^{-}_k
E^{-}_{k+q}}\right] \left(\frac{1}{E^{d}_{k} - E^{d}_{k+q} -
\omega} + \frac{1}{E^{d}_{k} - E^{d}_{k+q} +
\omega}\right)
\label{chizz}
\end{eqnarray}
(upper sign for $\chi^{zz}$ and lower for $\chi^{\rho\rho}$), then
\begin{eqnarray}
\chi^{z\rho}_{q,-q} &=&
 \frac{1}{8N} ~\sum_{E^{d}_k <\mu}~ \left[
\frac{\epsilon^{-}_k}{E^{-}_k} -
\frac{\epsilon^{-}_{k+q}}{E^{-}_{k+q}}\right]
 \left(\frac{1}{E^{c}_{k+q} -
 E^{d}_{k} - \omega} -
\frac{1}{ E^{c}_{k+q} - E^{d}_k + \omega}\right) +
 \nonumber \\
&&  -\frac{1}{8N} \sum_{\stackrel{E^{d}_{k+q} <\mu}{E^{d}_{k} >\mu}}
\left[\frac{\epsilon^{-}_{k}}{E^{-}_{k}} +
\frac{\epsilon^{-}_{k+q}}{E^{-}_{k+q}}\right]
\left( \frac{1}{E^{d}_{k} - E^{d}_{k+q} -
\omega} -  \frac{1}{E^{d}_{k} - E^{d}_{k+q} +
\omega} \right) .
\label{chizrho}
\end{eqnarray}
Further,
\begin{eqnarray}
\chi^{+-}_{q,-q} &=&
 \frac{1}{4N} \sum_{E^{d}_{k-{\bar Q}} <\mu}~ \left[1 +
\frac{\epsilon^{-}_{k-{\bar Q}} \epsilon^{-}_{k+q}}{E^{-}_{k-{\bar Q}}
E^{-}_{k+q}} - \frac{\epsilon^{-}_{k-{\bar Q}}}{E^{-}_{k-{\bar Q}}}
- \frac{\epsilon^{-}_{k+q}}{E^{-}_{k+q}} \right]
 \left(\frac{1}{E^{c}_{k+q} -
 E^{d}_{k-{\bar Q}} - \omega} +  \frac{1}{E^{c}_{k+q} -
 E^{d}_{k-{\bar Q}}
 + \omega} \right) +
\nonumber \\
&&  \frac{1}{4N}~ \sum_{\stackrel{E^{d}_{k+q} <\mu}{E^{d}_{k-{\bar Q}} >\mu}}
 \left[1 -
\frac{\epsilon^{-}_{k-{\bar Q}} \epsilon^{-}_{k+q}}{E^{-}_{k-{\bar Q}}
E^{-}_{k+q}} - \frac{\epsilon^{-}_{k-{\bar Q}}}{E^{-}_{k-{\bar Q}}} +
 \frac{\epsilon^{-}_{k+q}}{E^{-}_{k+q}} \right]
 \left(\frac{1}{E^{d}_{k-{\bar Q}} - E^{d}_{k+q} -
\omega} + \frac{1}{E^{d}_{k-{\bar Q}} - E^{d}_{k+q} +
\omega}\right) ,
\label{chi+-}
\end{eqnarray}
and
\begin{eqnarray}
\chi^{++}_{q,-(q+2{\bar Q})} =
 -\frac{1}{4N} \sum_{E^{d}_{k-{\bar Q}} <\mu}
\frac{\Delta^2}{E^{-}_{k-{\bar Q}}
E^{-}_{k+q}} \left(\frac{1}{E^{c}_{k+q} -
 E^{d}_{k-{\bar Q}} - \omega} + \frac{1}{E^{c}_{k+q} -
 E^{d}_{k-{\bar Q}} + \omega}\right) + \nonumber \\
\frac{1}{4N}  \sum_{\stackrel{E^{d}_{k+q} <\mu}{E^{d}_{k-{\bar Q}} >\mu}}
\frac{\Delta^2}{E^{-}_{k-{\bar Q}}
E^{-}_{k+q}} \left(\frac{1}{E^{d}_{k-{\bar Q}} - E^{d}_{k+q} -
\omega} + \frac{1}{E^{d}_{k-{\bar Q}} - E^{d}_{k+q} +
\omega}\right) .
\label{chi++}
\end{eqnarray}
Next,
\begin{eqnarray}
&&\chi^{+z,+\rho}_{q,-(q+{\bar Q})} =
 \frac{\Delta}{8N} ~\sum_{E^{d}_{k-{\bar Q}} <\mu}~\left[
\frac{(E^{-}_{k+q} \pm E^{-}_{k-{\bar Q}}) - (\epsilon^{-}_{k+q} \pm
\epsilon^{-}_{k - {\bar Q}})}{E^{-}_{k+q} E^{-}_{k-{\bar Q}}}\right]
 \left(\frac{1}{E^{c}_{k+q} -
 E^{d}_{k-{\bar Q}} - \omega} \mp \frac{1}{E^{c}_{k+q} -
 E^{d}_{k-{\bar Q}} + \omega}\right) +
 \nonumber \\
&&  \frac{\Delta}{8N} \sum_{\stackrel{E^{d}_{k+q} <\mu}{E^{d}_{k-{\bar Q}}
>\mu}}
\left[\frac{(E^{-}_{k-{\bar Q}} \mp E^{-}_{k+q}) -
 (\epsilon^{-}_{k-{\bar Q}} \pm \epsilon^{-}_{k+q})}{E^{k-{\bar Q}}
E^{-}_{k+q}}\right]
\left( \frac{1}{E^{d}_{k-{\bar Q}} - E^{d}_{k+q} -
\omega} \mp  \frac{1}{E^{d}_{k-{\bar Q}} - E^{d}_{k+q} +
\omega}\right) .
\label{chi+zrho}
\end{eqnarray}
Again, upper sign is for $\chi^{+z}$, lower for $\chi^{+\rho}$.
Finally,
\begin{eqnarray}
&&\chi^{-z,-\rho}_{q,-(q+{\bar Q})} =
 -\frac{\Delta}{8N} ~\sum_{E^{d}_{k-{\bar Q}} <\mu}~\left[
\frac{(E^{-}_{k+{\bar Q}} \pm E^{-}_{k+q}) + (\epsilon^{-}_{k+{\bar Q}} \pm
\epsilon^{-}_{k +q})}{E^{-}_{k+q} E^{-}_{k+{\bar Q}}}\right]
 \left(\frac{1}{E^{c}_{k+q} -
 E^{d}_{k+{\bar Q}} - \omega} \mp \frac{1}{E^{c}_{k+q} -
 E^{d}_{k+{\bar Q}} + \omega}\right) +
 \nonumber \\
&&  \frac{\Delta}{8N} \sum_{\stackrel{E^{d}_{k+q} <\mu}{E^{d}_{k+{\bar Q}}
>\mu}}
\left[\frac{(E^{-}_{k+q} \mp E^{-}_{k+{\bar Q}}) -
 (\epsilon^{-}_{k+q} \pm \epsilon^{-}_{k+{\bar Q}})}{E^{k+{\bar Q}}
E^{-}_{k+q}}\right]
\left( \frac{1}{E^{d}_{k+{\bar Q}} - E^{d}_{k+q} -
\omega} \mp  \frac{1}{E^{d}_{k+{\bar Q}} - E^{d}_{k+q} +
\omega}\right) .
\label{chi-zrho}
\end{eqnarray}
We also have $ = \chi^{--}_{q+2{\bar Q},-q} = \chi^{++}_{q,-(q+2{\bar Q})},
\chi^{z+,\rho+}_{q+{\bar Q},-(q+2{\bar Q})}
= \chi^{-z,-\rho}_{q+2{\bar Q},-(q+{\bar Q})},~
\chi^{+z,+\rho}_{q,-(q+{\bar Q})}
= \chi^{z-,\rho-}_{q+{\bar Q},-q}$.

\section{appendix B}
\label{append B}

In this appendix, we show that the zero modes in the transverse
susceptibilities
of the two spiral phases exist, to order $x^2$,
independent of the ratio of $t/U$ and $t^{\prime}/U$.
To see this, consider the transverse susceptibility right at $q = (\pi,\pi)$.
We will show that $\chi^{zz} = 1/2U$, i.e., the
 total ${\overline \chi}^{zz}_{\pi,-\pi} = \chi^{zz} ( 1 - 2U \chi^{zz})^{-1}$
 diverges despite the fact that for the spiral states, $(\pi,\pi)$ is not the
ordering momentum.  For definiteness, we will perform the calculations for
the $(\pi,Q)$ phase.
The calculations for the $(Q,Q)$ phase proceed in the same way, and the final
result is valid for both spiral states.

Expanding in (\ref{chizz}) to the second order in ${\bar q} \sim x$ we obtain
\begin{eqnarray}
\chi^{zz}_{\pi,-\pi} &=&
 \frac{1}{2N} \sum_{E^{d}_k <\mu}~
\frac{1}{E^{c}_{k+\pi} -
 E^{d}_k} - \frac{1}{8N} \sum_{E^{d}_k <\mu}~
(\epsilon^{-}_k + \epsilon^{-}_{k+\pi})^2 ~\frac{\Delta^2}{(E^{-}_k)^4}
{}~\frac{1}{E^{c}_{k+\pi} -
 E^{d}_k} + \nonumber \\
&&  \frac{1}{8N} \sum_{\stackrel{E^{d}_{k+\pi} <\mu}{E^{d}_{k} >\mu}}
(\epsilon^{-}_k + \epsilon^{-}_{k+\pi})^2 ~\frac{\Delta^2}{(E^{-}_k)^4}
{}~\frac{1}{E^{d}_{k} - E^{d}_{k+\pi}}
\label{b1}
\end{eqnarray}
where $\pi$ should be interpreted as 2D momentum $(\pi,\pi)$, and as before,
${\bar q} = \pi - Q$. It is also convenient to redefine the momenta
such that $\epsilon^{+-}_k = (\epsilon_{k +Q} \pm \epsilon_k)/2$.
For the $(\pi,Q)$ spiral, we then have
\begin{equation}
\epsilon^{-}_k + \epsilon^{-}_{k+\pi} = 4|t^{\prime}| ~\cos k_x
{}~\sin k_y ~{\bar q} ,
\label{b2}
\end{equation}
and also
\begin{equation}
\epsilon^{+}_{k+\pi} - \epsilon^{+}_{k} = -2t ~{\bar q}~ \sin k_y .
\label{b3}
\end{equation}
Substituting (\ref{b2}) and (\ref{b3}) into (\ref{b1}), we find after
some simple algebra
\begin{equation}
\chi^{zz}_{\pi,-\pi} = \frac{1}{2U} + \frac{t^2 {\bar q}^2}{4N} \sum_{k}
\frac{(\sin k_y)^2}{(E^{-}_k)^3} - \frac{(t^{\prime})^2~{\bar q}^2}{N}
 ~\sum_k \frac{(cos k_x \sin k_y)^2}{(E^{-}_k)^3}
 - \frac{t {\bar q} x}{4 \delta^2} .
\label{b4}
\end{equation}
In obtaining this result, we used the relation
\begin{equation}
\frac{1}{2N} \sum_{k} \left(\frac{1}{E^{-}_{k+\pi} +
 E^{-}_k} - \frac{1}{2 E^{-}_k}\right) =  - 2 (t^{\prime})^2 {\bar q}^2
 \frac{1}{N} \sum_{k}~ \frac{(sin k_y \cos k_x)^2
(\epsilon^{-}_k)^2}{(E^{-}_k)^5}
\label{b6}
\end{equation}
which can be derived by straightforward computations using (\ref{b2}).
In (\ref{b4}) and (\ref{b6}), the summation is over the whole Brillouin zone.

Notice that the pocket contribution (a term $\sim t {\bar q} x$ in
(\ref{b4})) is the same as in the analysis
in the bulk of the paper, where we assumed that $t^{\prime}, t \ll U$. This
is simply related to the fact that the pockets are located at
$(\pi/2, \pm \pi/2)$ where both $\epsilon^{+}_k \text{and} \epsilon^{-}_k$ are
small compared to $\Delta$ independent of the ratio of the parameters.

We now need the exact relation between ${\bar q}$ and $x$, valid to the first
order in $x$, but for arbitrary $t,t^{\prime}$ and $U$. To find this relation,
we again compute the ground state energy of the $(\pi,Q)$ spiral, but this
time without assuming that $U$ is large compared to the hopping integrals.
Doing the same computations as in Sec.~\ref{m-f}, we find
\begin{eqnarray}
E^{(\pi,Q)} &=& -t {\bar q} x +
t^2{\bar q}^2 ~\frac{1}{N} \sum_{k} \left(\frac{2 \cos^{2} k_y -
\sin^{2} k_y}{2 E^{-}_k} + \frac{ \sin^{2} k_y
(\epsilon^{-}_{k})^2}{2(E^{-}_k)^3} \right) -\nonumber \\
&&  2 (t^{\prime})^2 \Delta^2
 {\bar q}^2 ~\frac{1}{N} \sum_{k}~
\frac{\sin^{2} k_y \cos^{2} k_x}{(E^{-}_k)^3} .
\label{b7}
\end{eqnarray}
The equilibrium ${\bar q}$  then satisfies
\begin{equation}
t {\bar q} x = t^2{\bar q}^2 ~\frac{1}{N} \sum_{k} \left(\frac{2 \cos^{2} k_y -
\sin^{2} k_y}{E^{-}_k} + \frac{ \sin^{2} k_y
(\epsilon^{-}_{k})^2}{(E^{-}_k)^3} \right) - 4(t^{\prime})^2 \Delta^2
 {\bar q}^2 ~\frac{1}{N} \sum_{k}~
\frac{\sin^{2} k_y \cos^{2} k_x}{(E^{-}_k)^3} .
\label{b8}
\end{equation}
Substituting this result into (\ref{b4}), we obtain
\begin{equation}
\chi^{zz}_{\pi,-\pi} = \frac{1}{2U} + \frac{1}{2}~ t^2 {\bar q}^2~\Lambda ,
\end{equation}
where
\begin{equation}
\Lambda = \frac{1}{N} \sum_{k} \frac{\Delta^2 \sin^{2} k_y}{(E^{-}_k)^3}
 - \frac{1}{N} \sum_{k} \frac{\cos^{2} k_y}{E^{-}_k} .
\label{b9}
\end{equation}
Notice that all terms with $t^{\prime}$ are cancelled out.
Finally, to evaluate $\chi^{zz}$ to order $x^2$,
we actually need $\Lambda$ only for ${\bar q}=0$.
In this case, $\epsilon^{-}_k = -2t (\cos k_x + \cos k_y)$, and
integrating by parts in (\ref{b9}), we  immediately
obtain that $\Lambda =0$ is independent of
the ratio of $t/U$. We thus find
\begin{equation}
\chi^{zz}_{\pi,-\pi} = \frac{1}{2U} + O(x^3)
\end{equation}
This result implies that the zero modes in  the transverse susceptibility exist
at an arbitrary ratio of the parameters of the Hubbard model provided that
 the magnetic ordering at half-filling is
commensurate, and  doped holes form pockets at $(\pi/2,\pi/2)$ and symmetry
related points.

\begin{figure}
\caption{Spin configuration of a noncoplanar state. Arrows with thick
ends point out of the plane, while those with thick tails --- into the plane.
This configuration is different from the double spiral considered in
\protect\cite{lee}b. }
\label{nonplanara}
\end{figure}

\begin{figure}
\caption{Two adjacent spins in the equilibrium configuration.
 The in-plane
component, $S_\perp \sim x^{1/2}$,
  is small compared to the off-plane component, $S_\parallel$.}
\label{nonplanarb}
\end{figure}

\begin{figure}
\caption{The out-of-plane static susceptibility for the planar $(\pi,Q)$
spiral. The susceptibility is negative around $(\pi,\pi)$ indicating
instability towards spontaneous magnetization in the out-of-plane direction}
\label{fig_chiinst}
\end{figure}

\begin{figure}
\caption{The in-plane static susceptibility for the equilibrium noncoplanar
state, and a noncoplanar state with vanishing $\Delta_\perp$. }
\label{fig_chi_pm}
\end{figure}

\begin{figure}
\caption{The out-of-plane static susceptibility for three noncoplanar states:
one with $\Delta_\perp \to 0$, another with an intermediate value of
$\Delta_\perp$, and the third with the equilibrium value of $\Delta_\perp$.}
\label{fig_chi_zz}
\end{figure}

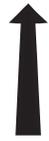
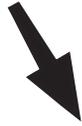
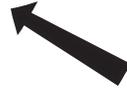
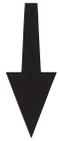
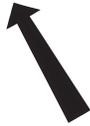
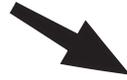
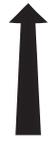
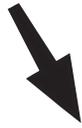
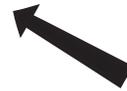
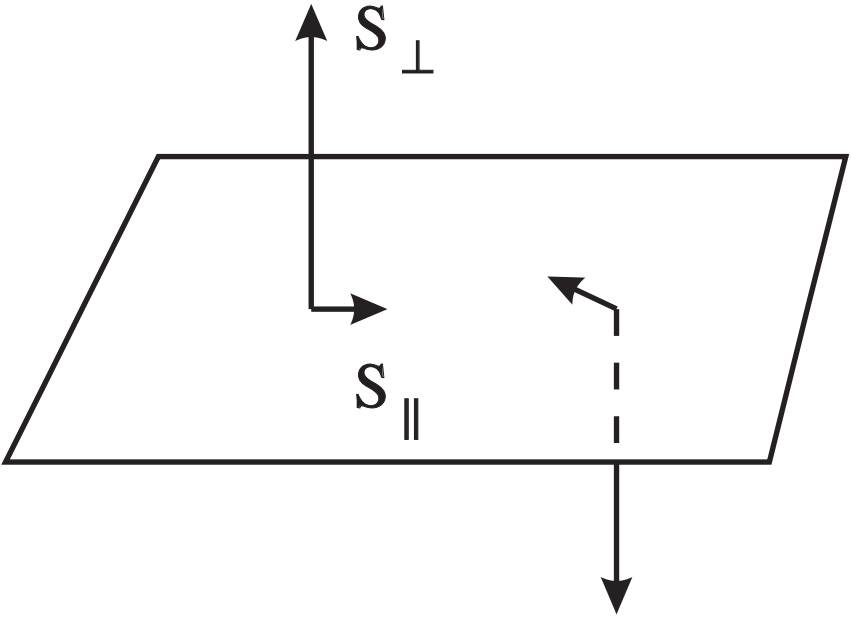
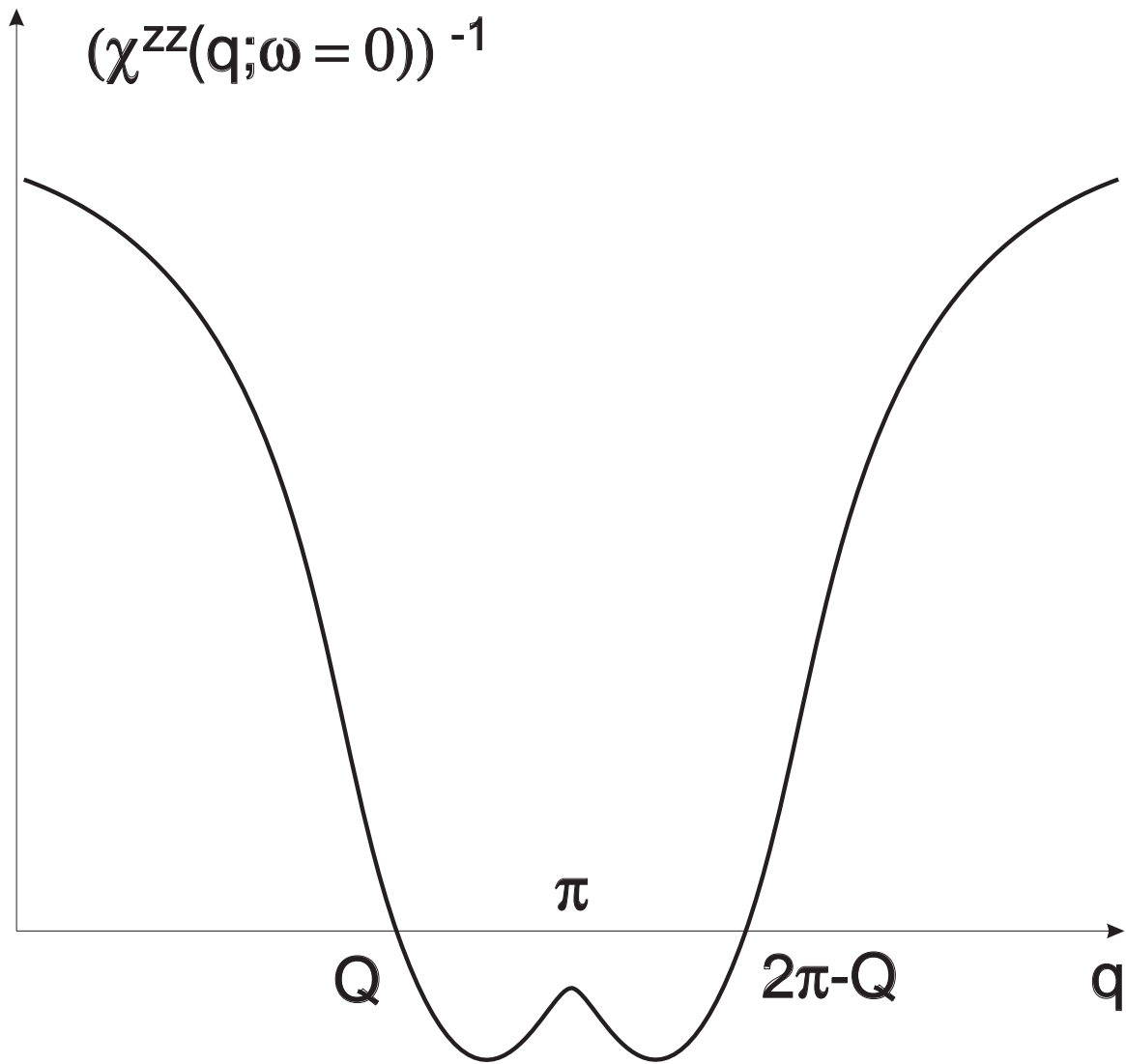
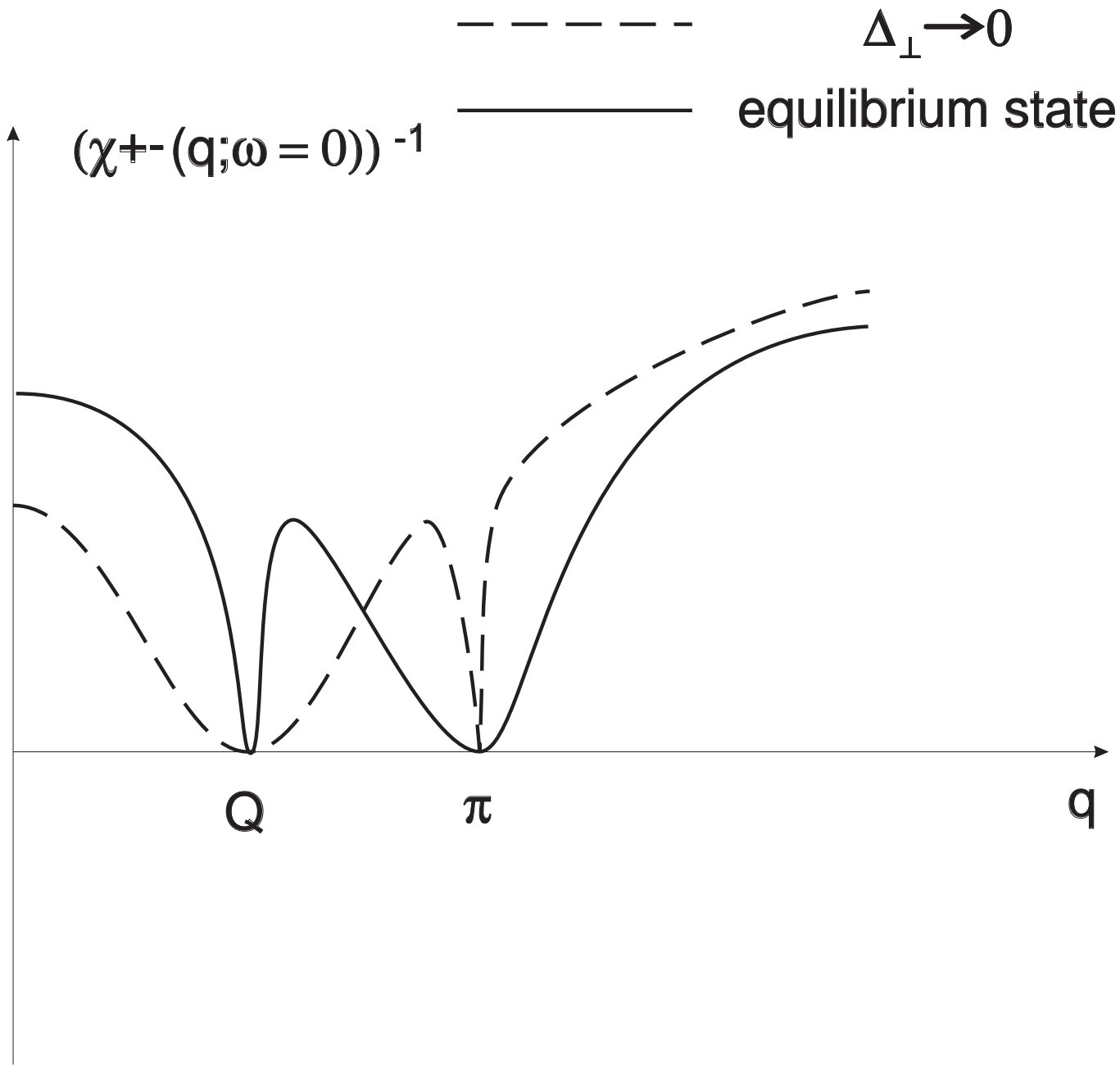
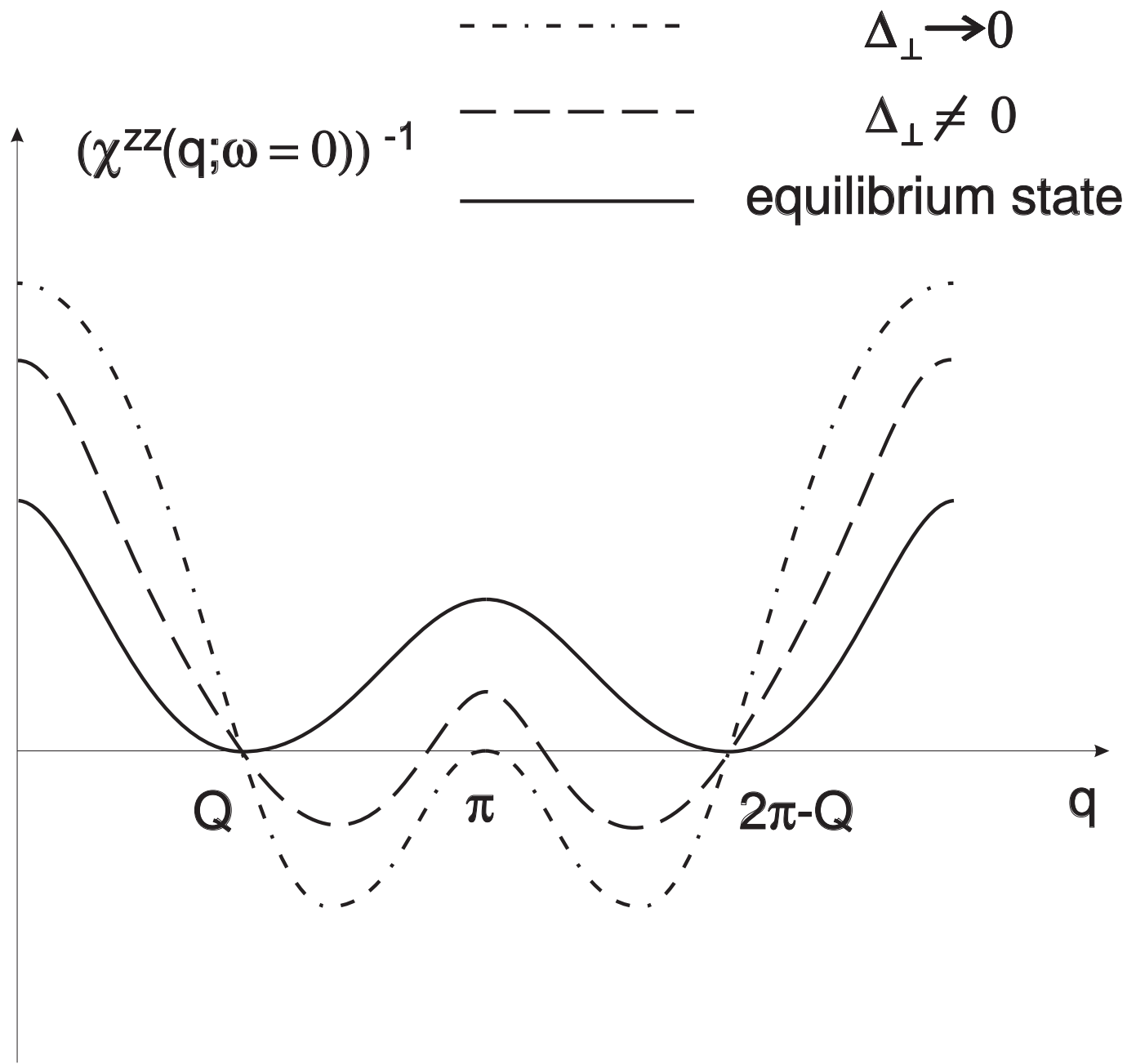
\end{document}